\documentclass[
 reprint,
superscriptaddress,
 amsmath,amssymb,
 aps,
floatfix,
]{revtex4-2}
\usepackage[utf8]{inputenc}
\usepackage{amssymb}
\usepackage{braket}
\usepackage{booktabs}
\usepackage{float}
\usepackage{physics}
\usepackage{amsmath}    
\usepackage{graphicx}   
\usepackage{verbatim}   
\usepackage{color}      
\usepackage{subfigure}  
\usepackage{hyperref}   
\usepackage{algorithm,algorithmic}
\usepackage{array}
\newcolumntype{P}[1]{>{\centering\arraybackslash}p{#1}}

\newcommand{\be}{\begin{equation}}
\newcommand{\ee}{\end{equation}}
\newcommand{\ba}{\begin{eqnarray}}
\newcommand{\ea}{\end{eqnarray}}

\usepackage{longtable}

\usepackage{graphicx}
\usepackage{dcolumn}
\usepackage{bm}
\usepackage[mathlines]{lineno}

\begin{document}

\preprint{APS/123-QED}

\title{
Refined topology of the $N = 20$ island of inversion with high precision mass measurements of $^{31-33}$Na and $^{31-35}$Mg 
}

\author{E. M. Lykiardopoulou}
     \affiliation{TRIUMF, 4004 Wesbrook Mall, Vancouver, BC V6T 2A3, Canada}
     \affiliation{Department of Physics and Astronomy, University of British Columbia, Vancouver, British Columbia, V6T 1Z1, Canada}    
\author{C. Walls}
    \affiliation{TRIUMF, 4004 Wesbrook Mall, Vancouver, BC V6T 2A3, Canada}
    \affiliation{Department of Physics and Astronomy, University of Manitoba, Winnipeg, Manitoba R3T 2N2, Canada}
\author{J. Bergmann}
    \affiliation{II. Physikalisches Institut, Justus Liebig University Gie\ss{}en, 35392 Gie\ss{}en, Germany}
\author{M. Brodeur}
    \affiliation{Department of Physics and Astronomy, University of Notre Dame, Notre Dame, Indiana, 46556, United States}
\author{C. Brown}
    \affiliation{School of Physics and Astronomy, University of Edinburgh, Edinburgh, EH9 3FD, United Kingdom}
\author{J. Cardona}
    \affiliation{TRIUMF, 4004 Wesbrook Mall, Vancouver, BC V6T 2A3, Canada}
    \affiliation{Department of Physics and Astronomy, University of Manitoba, Winnipeg, Manitoba R3T 2N2, Canada}
\author{A. Czihaly}
    \affiliation{TRIUMF, 4004 Wesbrook Mall, Vancouver, BC V6T 2A3, Canada}
    \affiliation{Department of Physics and Astronomy, University of Victoria, Victoria, British Columbia, V8P 5C2, Canada}
\author{T. Dickel}
    \affiliation{GSI Helmholtzzentrum f\"{u}r Schwerionenforschung GmbH, 64291 Darmstadt, Germany}
    \affiliation{II. Physikalisches Institut, Justus Liebig University Gie\ss{}en, 35392 Gie\ss{}en, Germany}
\author{T. Duguet}
    \affiliation{IRFU, CEA, Universit\'e Paris-Saclay, 91191 Gif-sur-Yvette, France}
    \affiliation{KU Leuven, Instituut voor Kern- en Stralingsfysica, 3001 Leuven, Belgium}
\author{J.-P. Ebran}
\affiliation{CEA, DAM, DIF, 91297 Arpajon, France}
    \affiliation{Universit\'e Paris-Saclay, CEA, Laboratoire Mati\`ere en Conditions Extr\^emes, 91680, Bruy\`eres-le-Ch\^atel, France}
\author{M. Frosini}
    \affiliation{CEA, DES, IRESNE, DER, SPRC, LEPh, 13108 Saint-Paul-l\`es-Durance, France}
\author{Z. Hockenbery}
     \affiliation{TRIUMF, 4004 Wesbrook Mall, Vancouver, BC V6T 2A3, Canada}
     \affiliation{Physics Department, McGill University, Montr\'eal, Qu\'ebec, H3A 2T8, Canada}
\author{J. D. Holt}
     \affiliation{TRIUMF, 4004 Wesbrook Mall, Vancouver, BC V6T 2A3, Canada}
     \affiliation{Physics Department, McGill University, Montr\'eal, Qu\'ebec, H3A 2T8, Canada}
\author{A. Jacobs}
     \affiliation{TRIUMF, 4004 Wesbrook Mall, Vancouver, BC V6T 2A3, Canada}
     \affiliation{Department of Physics and Astronomy, University of British Columbia, Vancouver, British Columbia, V6T 1Z1, Canada}
\author{S. Kakkar}
    \affiliation{TRIUMF, 4004 Wesbrook Mall, Vancouver, BC V6T 2A3, Canada}
    \affiliation{Department of Physics and Astronomy, University of Manitoba, Winnipeg, Manitoba R3T 2N2, Canada}
\author{B. Kootte}
    \affiliation{Department of Physics, University of Jyv\"{a}skyl\"{a}, Jyv\"{a}skyl\"{a}, Finland}
\author{T. Miyagi}
    \affiliation{TRIUMF, 4004 Wesbrook Mall, Vancouver, BC V6T 2A3, Canada}
    \affiliation{Technische Universit\"{a}t Darmstadt, Department of Physics, 64289 Darmstadt, Germany}
\author{A. Mollaebrahimi}
    \affiliation{TRIUMF, 4004 Wesbrook Mall, Vancouver, BC V6T 2A3, Canada}
    \affiliation{II. Physikalisches Institut, Justus Liebig University Gie\ss{}en, 35392 Gie\ss{}en, Germany}
\author{T. Murboeck}
    \affiliation{II. Physikalisches Institut, Justus Liebig University Gie\ss{}en, 35392 Gie\ss{}en, Germany}
\author{P. Navratil}
    \affiliation{TRIUMF, 4004 Wesbrook Mall, Vancouver, BC V6T 2A3, Canada}
    \affiliation{Department of Physics and Astronomy, University of Victoria, Victoria, British Columbia, V8P 5C2, Canada}
\author{T.~Otsuka}
    \affiliation{Department of Physics, The University of Tokyo, 7-3-1 Hongo, Bunkyo, Tokyo 113-0033, Japan}
    \affiliation{RIKEN Nishina Center, 2-1 Hirosawa, Wako, Saitama 351-0198, Japan}
\author{W. R. Pla\ss{}}
    \affiliation{II. Physikalisches Institut, Justus Liebig University Gie\ss{}en, 35392 Gie\ss{}en, Germany}
    \affiliation{GSI Helmholtzzentrum f\"{u}r Schwerionenforschung GmbH, 64291 Darmstadt, Germany}
\author{S. Paul}
     \affiliation{TRIUMF, 4004 Wesbrook Mall, Vancouver, BC V6T 2A3, Canada}
     \affiliation{Ruprecht-Karls-Universit\"{a}t Heidelberg, D-69117, Heidelberg, Germany}
\author{W. S. Porter}
     \affiliation{Department of Physics and Astronomy, University of Notre Dame, Notre Dame, Indiana, 46556, United States}
\author{M. P. Reiter}
    \affiliation{School of Physics and Astronomy, University of Edinburgh, Edinburgh, EH9 3FD, United Kingdom}
\author{A.~Scalesi}
    \affiliation{IRFU, CEA, Universit\'e Paris-Saclay, 91191 Gif-sur-Yvette, France}
\author{C. Scheidenberger}
    \affiliation{GSI Helmholtzzentrum f\"{u}r Schwerionenforschung GmbH, 64291 Darmstadt, Germany}
    \affiliation{II. Physikalisches Institut, Justus Liebig University Gie\ss{}en, 35392 Gie\ss{}en, Germany}
    \affiliation{Helmholtz Research Academy Hesse for FAIR (HFHF), GSI Helmholtz Center
for Heavy Ion Research, Campus Gie\ss{}en, 35392 Gie\ss{}en, Germany}
\author{V. Som\`a}
    \affiliation{IRFU, CEA, Universit\'e Paris-Saclay, 91191 Gif-sur-Yvette, France}
\author{N. Shimizu}
    \affiliation{Center for Computational Sciences, University of Tsukuba, 1-1-1 Tennodai, Tsukuba, Ibaraki 305-8577, Japan}
\author{Y. Wang}
     \affiliation{TRIUMF, 4004 Wesbrook Mall, Vancouver, BC V6T 2A3, Canada}
     \affiliation{Department of Physics and Astronomy, University of British Columbia, Vancouver, British Columbia, V6T 1Z1, Canada}
\author{D. Lunney}
    \affiliation{Universit\'e Paris-Saclay, CNRS/IN2P3, IJCLab, 91405 Orsay, France}
\author{J. Dilling}
    \affiliation{TRIUMF, 4004 Wesbrook Mall, Vancouver, BC V6T 2A3, Canada}
     \affiliation{Department of Physics and Astronomy, University of British Columbia, Vancouver, British Columbia, V6T 1Z1, Canada}
     \affiliation{Duke University, Durham, NC 27708, USA}
     \affiliation{Oak Ridge National Laboratory, Oak Ridge, TN 37830, USA}
\author{A.A. Kwiatkowski}
    \affiliation{TRIUMF, 4004 Wesbrook Mall, Vancouver, BC V6T 2A3, Canada}
    \affiliation{Department of Physics and Astronomy, University of Victoria, Victoria, British Columbia, V8P 5C2, Canada}

\begin{abstract}
Mass measurements of $^{31-33}$Na and $^{31-35}$Mg using the TITAN MR-TOF-MS at TRIUMF's ISAC facility are presented, with the uncertainty of the $^{33}$Na mass reduced by over two orders of magnitude.  
The excellent performance of the MR-TOF-MS has also allowed the discovery of a millisecond isomer in $^{32}$Na.   
The precision obtained shows that the binding energy of the normally closed $N=20$ neutron shell reaches a minimum for $^{32}$Mg but \emph{increases} significantly for $^{31}$Na, hinting at the possibility of enhanced shell strength toward the unbound $^{28}$O.  We compare the results with new ab initio predictions that raise intriguing questions of nuclear structure beyond the dripline.


\end{abstract}

\maketitle

The evolution of shell structure in atomic nuclei gives us some of the most important clues for understanding the forces at work binding nuclear quantum systems.
The early formulation of the nuclear shell model by Goeppert Mayer and Jensen \cite{Mayer_shell_model, Jensen_shell_model}
was inferred from the so-called ``magic'' numbers (8, 20, 28, 50, 82, 126) that correspond to the filling of orbitals, inspired by the atomic system.  
These shell closures are now known to evolve for nuclei in which the imbalance of neutrons to protons becomes more extreme.  

The first such illustration came from the pioneering on-line spectrometry work of Thibault et al. 1975 \cite{PhysRevC.12.644ThibaultNa31}, who showed that the $N=20$ shell-closure fingerprint in $^{31}$Na disappeared, demonstrating the non-universality of this magic number on the mass landscape.  Nuclear theory calculations quickly followed, suggesting the existence of deformation-driving (so-called intruder) orbitals that the more numerous neutrons preferred to occupy. The phenomenon was found in neighboring isotopes, establishing what is now termed an island of inversion, as the spherical ground state configuration was supplanted by the deformed intruders.  

A plethora of experimental and theoretical work has continued with examination of the lighter (for example \cite{OtsukaPRL2001}) and the heavier shell closures (for a review see \cite{rmp2020}), notably the very recent study of the doubly magic $^{78}$Ni \cite{Taniuchi2019}.  While $^{78}$Ni has resisted the erosion of its $N=50$ shell, theoretical studies of neighboring isotopes (e.g., $^{79}$Cu \cite{Welker2020}) show signs of deformation, hinting that the phenomenon of inversion is more universal with the $N=20$ island merging with that of $N=28$ and $N=40$ merging with $N=50$ \cite{Nowacki2016}.  

In addition to theoretical interpretation, many experimental observables are required to characterize this interesting phenomenon.  A critical one is the nuclear binding energy, obtained from mass measurements.  This field has evolved dramatically since the initial $^{31}$Na measurements (see \cite{tcp2019} and literature cited within), now offering exquisite precision and sensitivity by combining the use of Penning traps \cite{BlaumPT} and multi-reflection time-of-flight (MR-TOF) \cite{WOLLNIK201338, PLA2013134, WOLF201282, SchuryMRTOF} devices.  


The binding energy also defines the limits of the existence of bound nuclear systems:  the drip lines.  The $N=20$ island of inversion is of particular interest because of its proximity to the neutron drip line, more so because of the puzzling trajectory of the oxygen isotopes and the so-called oxygen anomaly \cite{oxyanom}.  

In this Letter, we present MR-TOF mass measurements performed at TRIUMF's ISAC facility of $^{31-33}$Na and $^{31-35}$Mg, which cover a large area of the $N=20$ island of inversion.  The greatly reduced uncertainties of these mass values reveal a clear sign that the $N=20$ shell effect re-emerges towards the neutron drip line, just three protons below.  In the absence of mass data, we  examine this exciting possibility in light of new ab initio theoretical predictions.   

~\\
~\\
The neutron-rich Mg and Na isotopes were produced by impinging up to 19~$\mu$A of 480 MeV proton beam onto a UCx target. The Na isotopes were surface ionized while the Mg isotopes were laser ionized using the Ion Guide Laser Ion Source (IGLIS) and TRIUMF's Resonant Laser Ionization Source (TRILIS) \cite{trilis} in two measurement campaigns. These ions were electrostatically accelerated to 20 keV and separated from other contaminant ions using the high resolution mass separator. As the resolution needed to separate Mg from Na ions (e.g., $R=m/ \Delta m = 1880$ for $^{31}\textrm{Mg}^+$-$^{31}\textrm{Na}^+$) is comparable to the resolving power of the mass separator ($R=2000$), the two species were simultaneously delivered through the Isotope Separator and ACcelerator (ISAC)
and delivered to TRIUMF's Ion Trap for Atomic and Nuclear science (TITAN) for the mass measurements. The beam was continuously accumulated and cooled in a Radio-Frequency Quadrupole Cooler/Buncher (RFQCB) \cite{BRUNNER201232} that uses helium gas at a pressure of 10$^{-2}$ mbar for buffer gas cooling. After approximately 12 ms of accumulation (80 Hz cycles), the ions were sent to the
MR-TOF-MS \cite{Jesch2015} where they were first further cooled in the MR-ToF RFQ transport system \cite{MRtof_2021} and then injected into the ToF analyzer. In the analyzer, they were reflected between two electrostatic mirrors for 506-668 isochronous turns, depending on the mass number, and detected on a MagneToF detector (ETP MagneTOF$^{\textrm{TM}}$) upon extraction. This resulted in a separation corresponding to a mass resolving power of 350,000 after a Time-Resolved Calibration (TRC) \cite{PhysRevC.99.064313GiessenMRTOF} is performed. 

\begin{figure*}
\includegraphics[width=2\columnwidth]{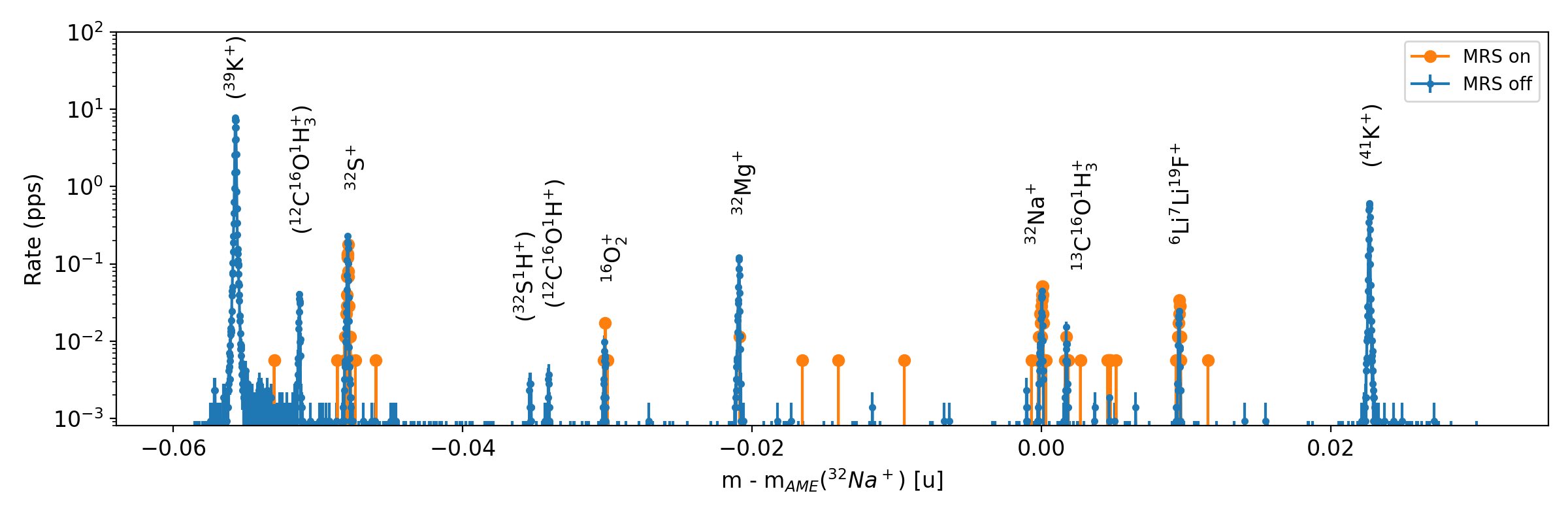}
\caption{\label{fig: mass 32 spectrum} Mass spectrum of the $A=32$ beam (blue) highlighting the purification achieved using the Mass Range Selector (MRS) kicker (orange). In the MRS off measurement (blue), non-isobaric species that perform a different number of isochronous turns are allowed in the trap. Non-isobaric species are marked with parentheses. The MRS-on measurement was shorter and therefore contains less statistics. In addition, the high resolution mass separator was fine-tuned to deliver the largest Mg yield during the MRS-off measurement, resulting in an increased rate of $^{32}$Mg.}
\end{figure*}

Three calibration steps were used in the derivation of the final mass values: i) a TRC calibration that accounts for drifts, ii) a peak-shape calibration that determines the shape of the time-of-flight peaks and iii) a mass calibration that determines the final mass value using a calibrant of well-known mass. 
Mass calibration was performed solely using isobaric species while for peak-shape and a time-resolved calibration, non-isobaric stable potassium beam from the MR-TOF-MS ion source was used when the rate of isobaric beam contaminants was low. 
An example of a spectrum containing non-isobaric calibrants is shown for $A=32$ in Fig. \ref{fig: mass 32 spectrum}. 
Since, in this case, the rate of the ions of interest was comparable to that of other isobaric ions, $^{39, 41}\textrm{K}^+$ beam was injected from the MR-TOF-MS ion source. Non-isobaric species perform a different number of isochronous turns in the MR-TOF-MS and can be removed by switching the Mass Range Selector (MRS) \cite{DICKEL2015172}. Fig. \ref{fig: mass 32 spectrum} shows a comparison of the mass spectrum when the MRS is off (blue) and on (orange), determining which species are isobaric.

\begin{figure}
\includegraphics[width=.9\columnwidth]{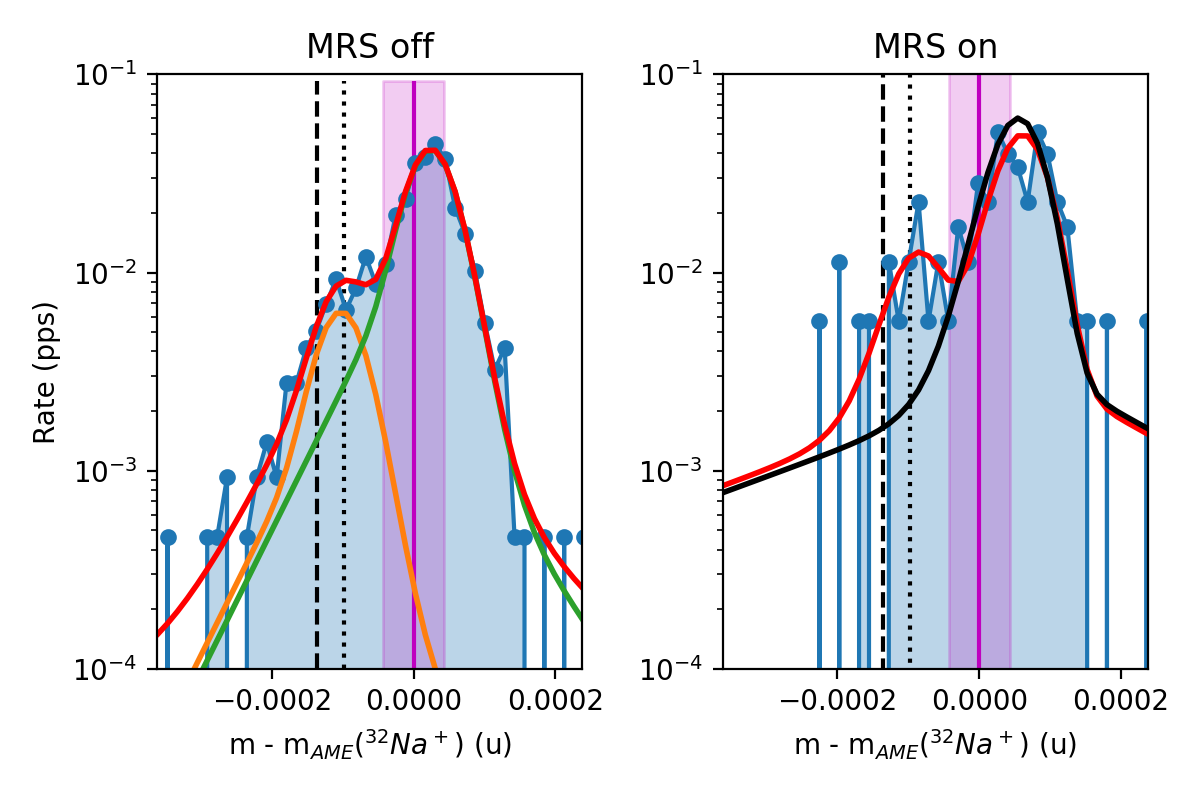}
\caption{\label{fig: Na32 spectrum} Mass spectrum of $^{32}$Na and $^{32}$Na$^m$ with the MRS off (left) and on (right). Shown in red, orange and green are the combined fit and the two single-peak fits respectively. The black line shows the best fit model under the hypothesis that there is only one peak with the MRS on. The purple vertical line shows the $^{32}$Na$^+$ literature mass along with its uncertainty (1$\sigma$). Vertical dashed and dotted lines show the candidate species for the first peak: $^2\textrm{H}^{7}\textrm{Li}^{23}\textrm{Na}^+$ and $^1\textrm{H}^{16}\textrm{O}_3^+$ which were rejected during analysis. For more information see main text.}
\end{figure}

\begin{figure}
\includegraphics[width=0.9\columnwidth]{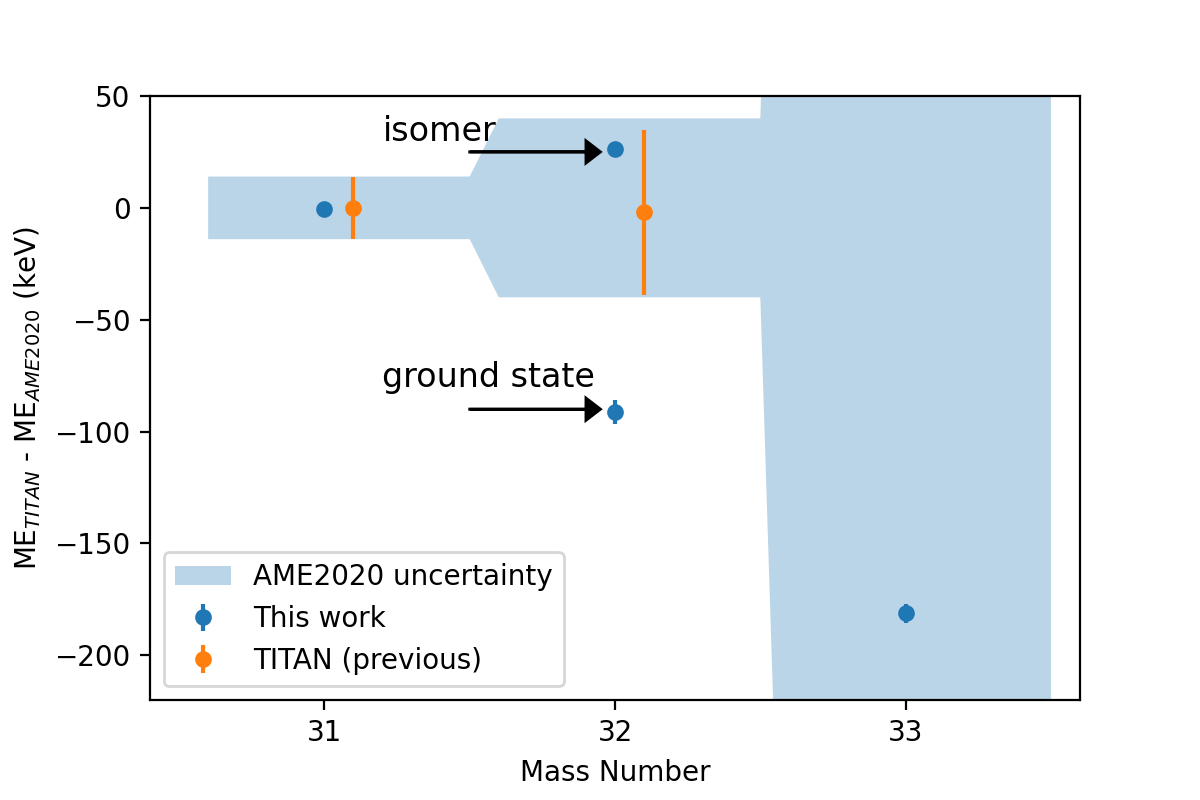}
\caption{\label{fig: Na deviation from AME} Deviation from literature for $^{31-33}$Na mass values. Previous TITAN measurements are shown in orange circles while the results of this work are shown in blue cicles. The AME mass uncertainty \cite{Wang_2021} is represented by a blue band. 
}
\end{figure}

The centroids of all peaks in each spectrum were determined using HyperEMG fit functions \cite{paul_stefan_f_2020_4731019emgfit}, an example of which is depicted in Fig. \ref{fig: Na32 spectrum}.
Systematic errors were evaluated and, depending on the measurement campaign and the number of ions stored in the MR-TOF-MS during each mass measurement, amount to a relative uncertainty of $8.4\cdot10^{-8} - 1.1\cdot10^{-7}$ which has been added quadratically with the statistical error for each mass number. For more details about the systematic uncertainty quantification see: \cite{Lykiardopoulou_2023}.
The mass excess values of the Mg and Na isotopes that were measured in the present work are shown in Table \ref{tab: Mg ME values}, along with their corresponding calibration masses. The AME2020 \cite{Wang_2021} mass excess values and uncertainties are also given for comparison.
A visual representation of the mass deviation from literature for the sodium isotopes is shown in Fig. \ref{fig: Na deviation from AME}. 

\begin{table*}[]
\centering
\begin{tabular}{P{2.5cm} P{2.5cm} P{2.5cm} P{2.5cm} P{2.5cm} P{2.5cm} P{2.5cm}}  
\hline \hline 
Species & Mass calibrant & Ionic mass ratio & ME (keV)  & ME$_{AME}$ (keV) & Difference (keV) \\  \hline  
$^{31}$Mg &  $^{12}\textrm{C}^{16}\textrm{O}^1\textrm{H}_3^{+}$  & 0.999298732(115)
 & -3131.8(3.9)  & -3122(3)            &     -9.8(5.0)              \\
$^{32}$Mg &  $^{32}$S$^{+}$ & 1.000845711(62) & -829.2(3.0)   &  -829(3)             &          -0.2(4.2)         \\
$^{33}$Mg &  $^1\textrm{H}^{16}\textrm{O}_2^{+}$ & 1.000232267(135) & 4954.1(3.6)   &  4963(3)             &          -8.9(4.7)        \\
$^{34}$Mg &  $^1\textrm{H}_2^{16}\textrm{O}_2^{+}$ & 1.000101321(118) & 8315.1(4.6)*   &  8323(7)             &        -7.9(8.4)           \\
$^{34}$Mg &  $^{10}B^{12}$C$_2^{+}$ & 0.999882252(131) &    &               &                 \\
$^{35}$Mg &  $^{35}$Cl$^{+}$ & 1.000209228(384) & 15529.5(7.1)*  &  15640(270)            &   -110(270) \\     
$^{35}$Mg &  $^{11}$B$^{12}$C$_2^{+}$ & 1.000210554(324) &   &              &    \\     
$^{31}$Na &  $^{12}\textrm{C}^{16}\textrm{O}^1\textrm{H}_3^{+}$ & 0.999830942(53) & 12245.3(2.6)  &  12246(14)            &      -1(14)            \\
$^{32}$Na &  $^{32}$S$^{+}$ & 1.001496385(167) &    18548.7(5.5)      & 18640(40)            &       -91(40)           \\
$^{32m}$Na &  $^{32}$S$^{+}$ & 1.001500334(76) &    18666.3(3.3)      & --            &         --        \\
$^{33}$Na &  $^1\textrm{H}^{16}\textrm{O}_2^{+}$ & 1.000838864(157) & 23598.8(4.3)  &  23780(450)            &  -180(450)
\\ \hline \hline 
\end{tabular}
\caption{\label{tab: Mg ME values} Mass Excess values of the isotopes measured in  this work. The last two columns correspond to the literature values from AME2020 \cite{Wang_2021}. 
The $^{32m}$Na isomer excitation energy is determined to be 117.6(6.4) keV. Mass excess values with asterisk are the weighted average of measurements that used different calibration species. Systematic errors have only added to the final Mass Excess values.}
\end{table*}

The mass uncertainties of $^{34,35}$Mg and $^{31-33}$Na were reduced by up to two orders of magnitude. We find that both $^{35}$Mg and $^{33}$Na, which are the most exotic isotopes measured, are more bound by more than 100 keV, despite being within 1$\sigma$ of their previously measured values \cite{JURADO200743Na33, Wang_2021}.

As shown in Fig. \ref{fig: Na32 spectrum}, a double peak appears in the mass vicinity of $^{32}\textrm{Na}^+$. A peak identification procedure was applied where the mass values of all relevant stable and radioactive ions and molecules (singly and doubly charged, with half-lives larger than 1 ms) were compared to the centroids of the two detected peaks. From this study, we find two species that align with the first peak. 
The first is $^2\textrm{H}^{7}\textrm{Li}^{23}\textrm{Na}^+$ which is 35 keV lighter than the detected peak (its expected position is shown in Fig. \ref{fig: Na32 spectrum} with a dashed line). We reject this candidate as it is over 4.5$\sigma$ lighter. The second candidate is the non-isobaric $^1\textrm{H}^{16}\textrm{O}_3^+$ from the MR-TOF-MS ion source (its expected position is shown in Fig. \ref{fig: Na32 spectrum} with a dotted line, assuming that it undergoes 530 isochronous turns).
To test whether the peak under discussion corresponds to the non-isobaric $^1\textrm{H}^{16}\textrm{O}_3^+$,  spectra with MRS off and on in Fig. \ref{fig: Na32 spectrum} are compared. Although less intense, both peaks appear in the MRS-on spectrum as well. Due to the low statistics, hypothesis testing was used to explore whether there are one or two peaks. The null hypothesis assumes one peak in the spectrum (black fit line in Fig. \ref{fig: Na32 spectrum}) while the alternative hypothesis assumes two peaks (red line in Fig. \ref{fig: Na32 spectrum}). Using the Monte Carlo and the Gross-Vitell method \cite{GVtest} to determine the $p$-value of the distribution, the alternative hypothesis is confirmed with a 4$\sigma$ confidence level, confirming both peaks to be isobaric.

Therefore, we conclude that both peaks correspond to $^{32}\textrm{Na}^+$, with the one on the left side of Fig. \ref{fig: Na32 spectrum} being the ground state and the one on the right side being a previously undiscovered low-lying isomeric state of $^{32}$Na. From the mass/energy difference between the two centroids, the excitation energy of the $^{32}$Na isomer is determined to be 117.6 (6.4) keV.

The mass of $^{32}$Na was previously measured using the TITAN Penning trap \cite{Galland2017}, the result of which (shown in Fig. \ref{fig: Na deviation from AME} in orange) is closer to $^{32m}$Na than to the ground state of $^{32}$Na (both shown in blue). Due to the short half-life of $^{32}$Na ($t_{1/2} = 13.2(4)$ ms \cite{ENSDF_32Na}), the excitation used in the Penning trap could not exceed 20~ms, limiting the resolving power to $\simeq 36,000$ (see \cite{Galland2017} for the description of the method and its limitation by the cyclotron frequency excitation period).  This resolving power is more than six times smaller than the resolving power required to separate two species of mass 32u that are only 118 keV apart, explaining why only one species was identified in \cite{Galland2017}. Furthermore, the weighted average of the mass excess of our measured $^{32}$Na and $^{32m}$Na agrees within one sigma with the measurement from \cite{Galland2017}, further supporting the hypothesis that the previous measurement is an average of the two states.

Very recently, Gray et al. \cite{Gray} reported the presence of a microsecond isomer in $^{32}$Na - the very first such state in the $N=20$ island of inversion.  Since their gamma-ray data were only shown until 150 $\mu$s after implantation, a millisecond isomer would not have been visible. 
While we did not determine the half-life of the new longer lived isomeric state, it is comparable to the ground state with a lower limit of a few milliseconds. Although theoretical approaches do predict a low-lying isomer for $^{32}$Na \cite{Gray}, a dedicated spectroscopy experiment or a half-life measurement in the MR-ToF MS (such as in \cite{Mukul_2020}) will be required to find out if the isomer that we found is the same as the one predicted by theory. 

\begin{figure}
\includegraphics[width=\columnwidth]{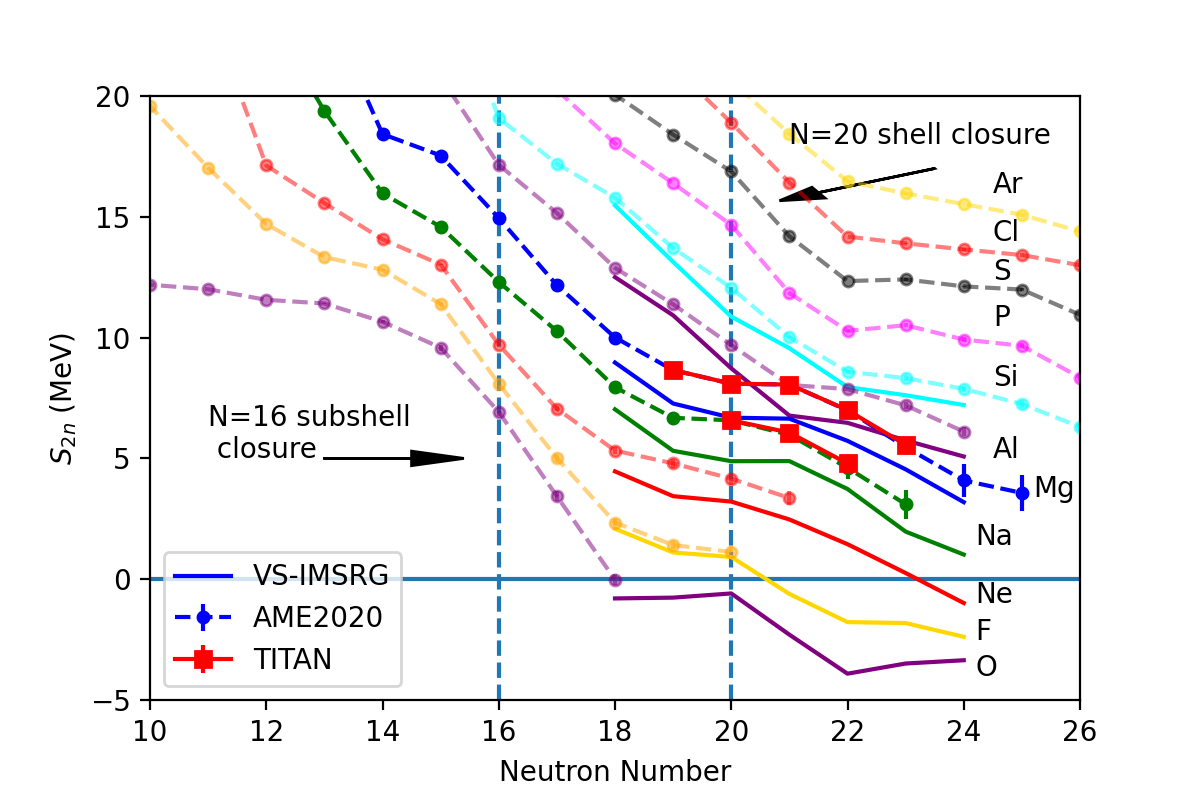}
\caption{\label{fig: theory S2n} 
Experimental two-neutron separation energies (from \cite{Wang_2021}) as a function of neutron number ($N$) for the isotopic chains  oxygen ($Z=8$) to argon ($Z=18$) with the new results (red symbols) and ab initio VS-IMSRG predictions for $Z=8-14$  from \cite{MiyagiIOI2020} (solid lines) superimposed.  
}
\end{figure}

~\\
~\\
We use the new results to obtain two-neutron separation energies, $S_{2n}=BE(Z,N)-BE(Z,N-2)$, where $BE$ is the binding energy obtained from the mass, which are shown in Fig.~\ref{fig: theory S2n} for $Z=8-18$.  The effect of a shell closure at $N=20$ -- a change in slope at $N=20$ and $N=22$ -- can be seen for sulphur ($Z=16$ - grey curve).  This sharp change smooths out for the magnesium and sodium isotopes showing how the shell strength is reduced, with a new shell closure appearing for oxygen ($Z=8$ - purple curve) at $N=16$. 

\begin{figure}
\includegraphics[width=\columnwidth]{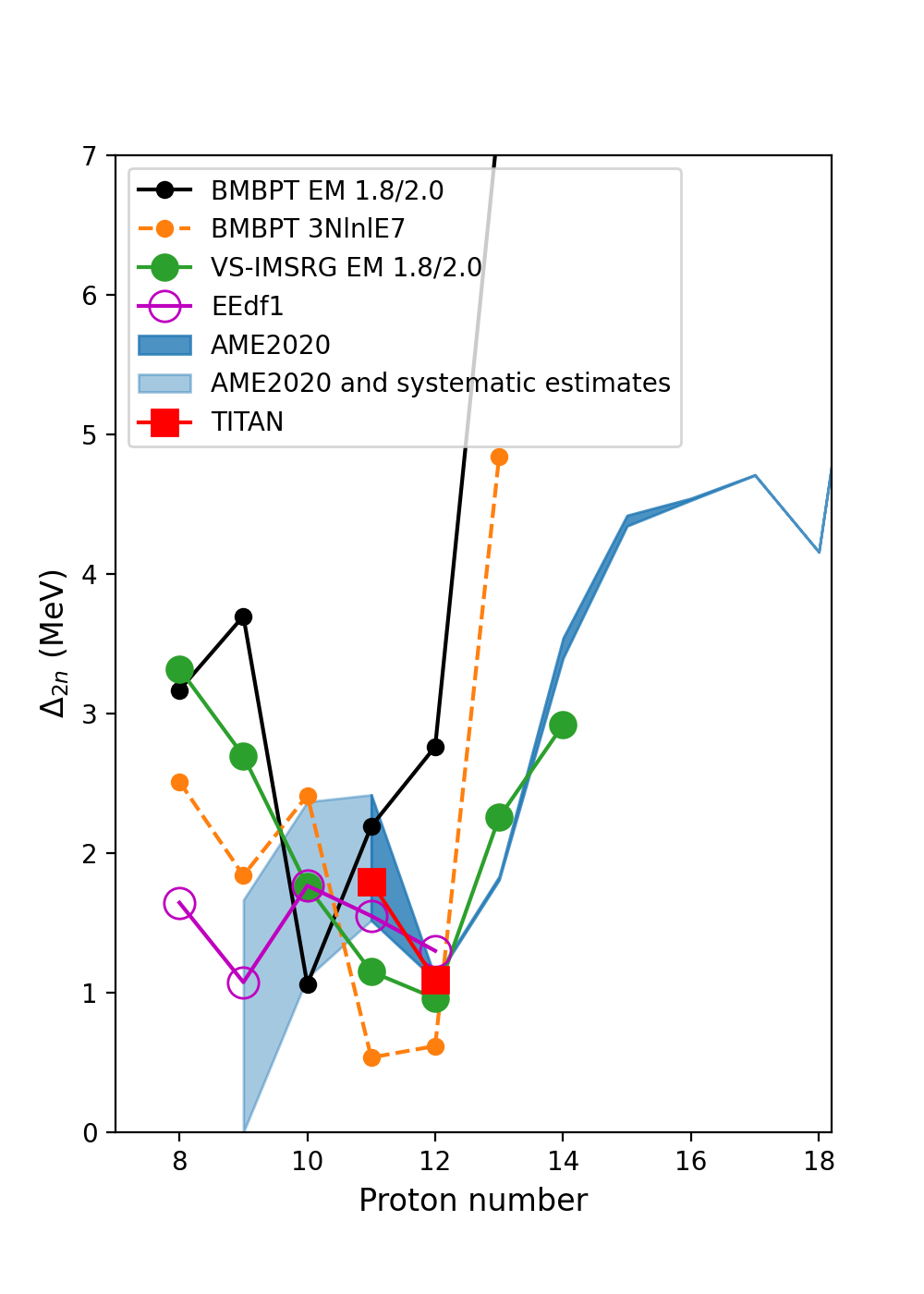}
\caption{\label{fig: Shell gap plot} The empirical shell gap for the $N=20$ isotones. The blue band represents AME2020 values and their 1$\sigma$ uncertainties \cite{Wang_2021} plus estimates, while red squares show the results of this work (uncertainties are smaller than the marker size). 
Green points depict the results of ab initio calculations using the VS-IMSRG \cite{MiyagiIOI2020} framework (derived from the curves in Fig. \ref{fig: theory S2n}).  Purple circles are the EEdf1 calculations published in \cite{Tsunoda2020}.  Orange and black depict ab initio theory calculations using the 3NLNLE7 interaction~\cite{Kravvaris2023} and the EM 1.8/2.0 interaction in the BMBPT \cite{BMBPT2} framework.}
\end{figure}

The effects are difficult to see at the 10 MeV scale of Fig.~\ref{fig: theory S2n} but a derivative of this surface enables closer examination of the relative shell strength using the
empirical shell gap, defined as the difference of $S_{2n}$ at and two neutrons beyond the shell closure:  $\Delta_{2n} = S_{2n}(Z,N)-S_{2n}(Z,N+2)$, shown in Fig.~\ref{fig: Shell gap plot} for $N=20$ isotones.

%
%
The stable isotope $^{36}$S ($Z=16$) has a ``nominal'' shell strength of 4-5 MeV.
Moving towards the dripline at lower $Z$, this shell strength is ``quenched'' due to deformation in the island of inversion.  The question of whether this strength is maintained, or disappears at the dripline is an important question for nuclear structure.  
The new results (red points in Fig.~\ref{fig: Shell gap plot})
clearly establish a minimum for $Z=12$ as the empirical shell strength \emph{increases} for $Z=11$.  This significant increase, from 1.1 to 1.8 MeV, hints at the enticing possibility that the $N=20$ shell closure could recover towards the drip-line oxygen isotopes. 

In the absence of experimental data, we must examine theoretical predictions.
First we use the multi-shell variant~\cite{MiyagiIOI2020} of the ab initio VS-IMSRG method (Valence-Space In-Medium Similarity Renormalization Group, described in \cite{Stro17ENO,Stro19ARNPS}), which has been applied across the medium-mass region~\cite{Stro21Drip}, including the island of inversion between calcium and nickel~\cite{Silw22Cr,Port22Fe} to compare with our new results. The $S_{2n}$ predictions obtained using the EM~1.8/2.0 chiral effective field theory interaction \cite{EM18/20_PhysRevC.83.031301,Simo17SatFinNuc}, in large three-body spaces~\cite{Miya22Heavy}, are plotted in Fig.~\ref{fig: theory S2n} as solid lines, where a generally good agreement with experiment can be seen. Despite a slight offset, the VS-IMSRG results correctly predict the trends in the island of inversion including the cross-over of Mg and Al at $N=21$ \cite{Kwiatkowski_crossover}.  The VS-IMSRG results for the shell gap are compared with the new TITAN result in Fig.~\ref{fig: Shell gap plot}. They correctly predict the drop in $\Delta_{2n}$ correlated with the island of inversion.  Moreover they match the minimum observed for $Z=12$ and predict a gradual \emph{increase} of the empirical shell gap that exceeds 3 MeV for the unbound $^{28}$O. 

Fig.~\ref{fig: Shell gap plot} also features the first systematic application of a new ab initio method:  Bogoliubov Many-Body Perturbation Theory (BMBPT) \cite{BMBPT1}, which now takes into account deformation, in addition to pairing \cite{BMBPT2}.  For comparison, the same chiral interaction is used.  Although the results deviate more than VS-IMSRG, similar behavior is observed for fluorine and oxygen showing a significant empirical shell gap rising to more than 3 MeV.  Both VS-IMSRG and BMBPT approaches correctly predict that $^{26}$O and $^{28}$O are unbound (as shown in Fig. \ref{fig: theory S2n} for VS-IMSRG).

Also included in Fig.~\ref{fig: Shell gap plot} are the shell model results from Tsunoda et al. \cite{Tsunoda2020} using the EEdf1 interaction, likewise derived from chiral effective field theory.  In addition to correctly predicting the dripline, these results are in excellent agreement with the new measurements presented here.  However, the predictions are drastically different for $Z=9$, indicating a strong quenching of the $N=20$ shell gap.  Although there are no mass data, the recent discovery by Ahn et al. \cite{Ahn2019} that $^{31}$F is bound, means that the corresponding shell gap for $Z=9$ must be lower than 1.5 MeV.  Considering that the $S_{2n}$ for $^{29}$F is 1.13(54) MeV \cite{Wang_2021} and assuming $S_{2n}$ = 0 for $^{31}$F, the EEdf1 prediction seems quite accurate.  
The EEdf1 calculations of \cite{Tsunoda2020} are extended to $^{30}$O here, so as to derive a shell gap for $Z=8$ in Fig.~\ref{fig: Shell gap plot}.  Although the value also \emph{increases}, it still remains relatively low.    


Fig.~\ref{fig: Shell gap plot} also includes first results achieved using a new chiral EFT interaction called NN(N$^4$LO)+3N$_{\rm lnlE7}$ \cite{Kravvaris2023} using the BMBPT ab initio method. This interaction combines the chiral NN 
potential from Ref.~\cite{Entem2017} with a 3N interaction 
regulated as discussed in Ref.~\cite{Soma2020} 
that includes a sub-leading contact term enhancing the spin-orbit strength 
~\cite{Girlanda2011}. The results are striking in their predictions resembling those of the successful EEdf1 with the same trend for Z = 8-10.




The ab initio predictions for the oxygen $N=20$ gap range from almost doubly magic (EM 1.8/2.0 interaction) to quenched (EEdf1) with the new NN(N$^4$LO)+3N$_{\rm lnlE7}$ result in between.  However, the energy range spans less than 2 MeV, which is a great step forward for these state-of-the-art approaches.  

Kondo et al. \cite{Kondo2023} recently reported the discovery of $^{28}$O,
obtaining crucial neutron binding energy data that are compared to theoretical predictions (see Fig.~3 in \cite{Kondo2023}), including the EEdf3 interaction (derived from EEdf1, cited above) and the same EM~1.8/2.0-based VS-IMSRG results.  All  approaches correctly predict $^{28}$O to be unbound (although they calculate a ground state energy that is 0.5 MeV - 2.25 MeV larger than what was measured experimentally), with the exception of the Gamow Shell Model \cite{GSM}, which predicts a (barely) bound $^{28}$O.  This model includes continuum effects, which can contribute sizeable binding energy components of 0.5 to 2.0 MeV.  None of the results in Fig.~\ref{fig: Shell gap plot} include such effects.  

Kondo et al. \cite{Kondo2023} also derived a spectroscopic factor for $^{28}$O revealing this unbound nuclide \emph{not} to be of doubly magic character, arguing for the extension of the island of inversion all the way to (unbound) oxygen.  However, their value lies between those of closed and open shell nuclides, indicating that other effects come into play.  

Spectroscopy of  $^{27,29}$F by Doornenbal et al. \cite{Doornenbal2017} confirmed an extension of the ``low-$Z$ shore'' of the island of inversion to $Z=9$, suggesting a ``persistent reduced neutron gap for $^{28}$O.''  While our shell gap for $^{31}$Na starts to increase towards the low-$Z$ shore, it could still be considered as ``reduced'' since it is still small.
Spectroscopy of the unbound isotopes $^{28}$F by Revel et al. \cite{Revel2020} and $^{30}$F by Kahlbow et al. \cite{KahlbowRIKEN} confirmed the island's extension to $Z=9$,
arguing against a doubly magic $^{28}$O.  

The question of shell strength beyond the \emph{proton} dripline was recently addressed by Lalanne et al. \cite{Lalanne2023}, who established $^{36}$Ca ($Z=20$) as a doubly magic nucleus from an enhanced $N=16$ shell gap, determined by a two-neutron transfer reaction $Q$-value.  Their shell gap (determined using $S_{n}$ values) is comparable to that of the doubly magic $^{24}$O, which is remarkable in that these nuclides lie at each end of the $N=16$ isotonic chain.  Moreover, the shell model calculations in \cite{Lalanne2023} predict a $N=14$ shell closure for $^{34}$Ca, which is unbound.  It is therefore not out of the question for unbound nuclides to reflect shell effects. 

Clearly, future theoretical developments will have to address the interplay of three-body and continuum effects, which unfortunately require enormous valence spaces.

Despite the new data on the unbound oxygen isotopes, the mass of $^{24}$O still has a relatively large uncertainty (160 keV).  This, as well as the more neutron-rich neon ($Z=10$) isotopes, will be the aim of the TITAN mass measurement program in the near future.  


~\\
~\\
In summary, we have performed the first precision mass measurements of $^{33}$Na and $^{35}$Mg and we have improved the precision of $^{31,32}$Na and $^{34}$Mg.  Following the recent discovery of a microsecond isomer in $^{32}$Na at FRIB, we believe we have discovered a millisecond isomer in the same nucleus, only 118 keV above the ground state.  (A dedicated spectroscopy measurement would be required for confirmation.)  The new mass of $^{33}$Na refines the empirical $N=20$ shell towards the neutron dripline, and its significantly increased value hints at the interesting possibility of the existence of strengthened shell behavior in the unbound oxygen isotopes.  
Ab initio results using new interactions and many body methods published here for the first time, show possible shell stabilization but do not yet agree on its extent.  
Precision masses are now clearly necessary to help ab initio theory untangle the interplay of the various effects that define the limits of nuclear stability.

\section{Acknowledgements}

This work was supported by the Natural Sciences and Engineering Research Council of Canada (NSERC), the National Research Council of Canada (NRC), by the French IN2P3, by the German Federal Ministry for Education and Research (BMBF)
under contracts no. 05P19RGFN1 and 05P21RGFN1, the German Research Foundation (DFG) under contract no. 422761894, and by Justus-Liebig-Universit\"{a}t Gie\ss{}en and GSI under the JLU-GSI strategic Helmholtz partnership agreement.
A.S. is supported by the European Union’s Horizon 2020 research and innovation program under grant agreement No 800945 - NUMERICS - H2020-MSCA-COFUND-2017. W.S.P. and M.B. acknowledge support from the US National Science Foundation under Grant No. PHY-2011890 and PHY-2310059.
BMBPT calculations were performed using HPC resources from GENCI-TGCC (Contract No. A0150513012).
T.O. and N.S. acknowledge the support of the ``Program for promoting research on the supercomputer Fugaku'', MEXT, Japan (JPMXP1020230411, hp230207, hp240213). In accordance with UKRI guidelines, the author has applied a Creative Commons Attribution (CC BY) licence to any Author Accepted Manuscript version arising from this submission.\\
~\\

\bibliography{apssamp}

\end{document}